\documentclass[numbers]{elsarticle}
\usepackage{latexsym}
\usepackage{float}
\usepackage[latin2]{inputenc}
\usepackage{graphicx}
\usepackage{ae}
\usepackage{aecompl}
\usepackage{setspace}
\usepackage{booktabs}
\usepackage{longtable}
\newtheorem{theorem}{Theorem}

\newtheorem{example}[theorem]{Example}

\usepackage[hidelinks]{hyperref}

\makeatletter
\def\ps@pprintTitle{%
	\let\@oddhead\@empty
	\let\@evenhead\@empty
	\def\@oddfoot{\centerline{\thepage}}%
	\let\@evenfoot\@oddfoot}
\makeatother

\title{Succinct Amyloid and Non-Amyloid Patterns in Hexapeptides}
		
		\author[p]{László Keresztes\corref{cor2}}
		\ead{keresztes@pitgroup.org}
		\author[p]{Evelin Szögi\corref{cor2}}
		\ead{szogi@pitgroup.org}
		\author[p]{Bálint Varga}
		\ead{balorkany@pitgroup.org}
		\author[s]{Viktor Farkas}
		\ead{farkasv@caesar.elte.hu}
		\author[s,t]{András Perczel}
		\ead{perczel@chem.elte.hu}
		\author[p,u]{Vince Grolmusz\corref{cor1}}
		\ead{grolmusz@pitgroup.org}
		\cortext[cor1]{Corresponding author}
		\cortext[cor2]{Joint first authors}
		\address[p]{PIT Bioinformatics Group, Eötvös University, H-1117 Budapest, Hungary}
		\address[u]{Uratim Ltd., H-1118 Budapest, Hungary}
		\address[s]{MTA-ELTE Protein Modeling Research Group, H-1117 Budapest, Hungary}
		\address[t]{Laboratory of Structural Chemistry and Biology, Eötvös University, H-1117, Budapest, Hungary}

\begin{document} 	

\begin{abstract}
Hexapeptides are widely applied as a model system for studying amyloid-forming properties of polypeptides, including proteins. Recently, large experimental databases have become publicly available with amyloidogenic labels. Using these datasets for training and testing purposes, one may build artificial intelligence (AI)-based classifiers for predicting the amyloid state of peptides. In our previous work (Biomolecules, 11(4) 500,  (2021)) we described the Support Vector Machine (SVM)-based Budapest Amyloid Predictor (\url{https://pitgroup.org/bap}). Here we apply the Budapest Amyloid Predictor for discovering numerous amyloidogenic and non-amyloidogenic hexapeptide patterns with accuracy between 80\% and 84\%, as surprising and succinct novel rules for further understanding the amyloid state of peptides. For example, we have shown that for any independently mutated residue (position marked by ``x''), the patterns CxFLWx, FxFLFx, or xxIVIV are predicted to be amyloidogenic, while those of PxDxxx, xxKxEx, and xxPQxx non-amyloidogenic at all. We note that each amyloidogenic pattern with two x's (e.g.,CxFLWx) describes succinctly $20^2=400$ hexapeptides, while the non-amyloidogenic patterns comprising four point mutations (e.g.,PxDxxx) gives $20^4=160,000$ hexapeptides in total. We also examine restricted substitutions for positions ``x'' from subclasses of proteinogenic amino acid residues; for example, if ``x'' is substituted with hydrophobic amino acids, then there exist patterns containing three x's, like MxVVxx, predicted to be amyloidogenic. If we can choose for the x positions  any hydrophobic amino acids, except the ``structure breaker'' proline, then we get amyloid patterns with five x positions, e.g., xxxFxx, each corresponding to 32,768 hexapeptides.  To our knowledge, no similar applications of artificial intelligence tools or succinct amyloid patterns were described before the present work. 
\end{abstract}

\date{}
	
\maketitle

\section*{Introduction} 

The amyloid formation of proteins and peptides has gained increasing attention in novel areas of medicine and biology in the last months, including the application of amyloidogenic aggregation cores in viral proteins as new antiviral agents \cite{Michiels2020}, or targeting the lethal transthyretin amyloidosis with human {\em in vivo} CRISPR-Cas9-based gene editing with high success rate \cite{Gillmore2021}. 

Amyloids are misfolded proteins \cite{Horvath2019,Taricska2020} with well-defined and periodic 3D structure, comprising mostly parallel $\beta$-sheets \cite{Takacs2019,Takacs2020}. While amyloids are only seldom present in healthy human tissues \cite{Maji2009}, they were reported to be connected with several neurodegenerative diseases \cite{Soto2006}, most importantly with Alzheimer's disease. 

In bioinformatics numerous amyloid-predictors were designed and published in the recent years, including APPNN \cite{Familia2015}, Zyggregator \cite{Tartaglia2008a}, AGGRESCAN \cite{Conchillo-Sole2007}, netCSSP \cite{Kim2009a}; for a recent review of their performance we refer to \cite{Santos2020}. These predictors use different machine learning approaches for decision making from training data. In our contribution \cite{Keresztes2020a} we applied a powerful but transparent machine learning tool, the linear Support Vector Machine (SVM) \cite{Cortes1995}. 

For building an SVM, we needed a training dataset of $n$-dimensional vectors $y^1,y^2,\ldots,y^m$, each labeled with a bit of either 0 or 1. From the training set, we determine a hyperplane, which, in a certain sense, ``optimally'' separates the 0 and the 1-labeled vectors in a way that most of the 1-labeled vectors are on one side of the hyperplane, and most of the 0-labeled ones are on the other side. 

The SVM-predictor now works as follows: If a new vector is situated on the 0-side, then the prediction is ``0'', and if it lies on the 1-side, the prediction is ``1''. 

Recently we built a Support Vector Machine for amyloidogenecity prediction of hexapeptides \cite{Keresztes2020a}. The training set was the Waltz dataset \cite{Beerten2015, Louros2020} of experimentally identified  514 amyloidogenic and 901 non-amyloidogenic hexapeptides. By applying the physico-chemical property dataset of amino acids,  AAindex \cite{Kawashima2008}, property vectors $y^1,y^2,\ldots,y^m$ were assigned to each hexapeptide, and using this multi-dimensional representation, we have prepared an SVM, called Budapest Amyloid Predictor (abbreviated as BAP), described in detail in \cite{Keresztes2020a}, and freely available at the address \url{https://pitgroup.org/bap}. 

The accuracy rate of our predictor is 84\%, more exactly, ACC = 0.84, TPR=0.75, TNR=0.9, PPV=0.8, NPV=0.86, (that is, accuracy, true positive ratio, true negative ratio, positive predictive value, negative predictive value, resp.). We remark that the accuracy of our SVM is better than or on par with that of APPNN \cite{Familia2015}, with a simpler, more transparent structure.

In the present contribution, we make use of this transparent structure of the Budapest Amyloid Predictor: we present numerous patterns related to amyloidicity, such that each of those patterns grasps hundreds or even tens of thousands of individual hexapeptides and give predictions of their amyloid-forming properties. For example, we show that for all (independent) substitutions of the 20 amino acids for letter ``x'', the hexapeptides  CxFLFx, FxFLWx, or xxIVIV are all predicted to be amyloids by the Budapest Amyloid Predictor. Note that each of these patterns describes $20^2=400$ different hexapeptides. We also note that no amyloid-forming patterns exist with three x's for the predictor. All the 5531 amyloid-forming hexapeptide patterns with two x's are listed in Table S1 in the Supporting Material.

We also show several patterns, which -- by the Budapest Amyloid Predictor -- would not form amyloids. For example, the patterns xxDDxx, xxPxDx, or xxPKxx with any (independently chosen) amino acids for the positions denoted by x are predicted to be non-amyloids by our Budapest Amyloid Predictor. Note that each of these patterns succinctly describes  $20^4=160,000$ hexapeptides. We add that non-amyloid patterns with five $x$ positions do not exist for our tool at \url{https://pitgroup.org/bap}. All the non-amyloid forming hexapeptide patterns with four $x$ positions are listed in Table 2 of the present contribution. 

We note that these patterns are succinct representations of the predictions of the Budapest Amyloid Predictor (BAP), whose accuracy rate is 84\% \cite{Keresztes2020a}; that is, we do not state, for example, that all CxFLFx hexapeptides are amyloids, but we state that all of them are predicted to be amyloids by the BAP tool.  The transparent linear structure of the Support Vector Machines makes possible the derivation of these intuitive, useful, and well-applicable patterns from an artificial intelligence (AI) tool, as we clarify in the present contribution.

We need to add also that today we are living in the era of the fast-developing AI methods and tools in numerous fields of science and technology. Most of these tools work as follows:

Suppose the tool needs to compute a value $f(y)$ from another value $y$. For constructing such an AI tool, the following steps are applied:

\begin{itemize}
\item A large set of previously acquired, correct $(y,f(y))$ pairs are partitioned into two classes: the training set A and the test set B.

\item The training set A is applied to construct a tool, which assigns the predicted value of $f(y)$, denoted by $f_p(y)$, to each $y$;

\item The test set B is used for evaluating the correctness of the tool: the predicted value, produced by the tool, $f_p(y)$ is compared to the correct, previously known $f(y)$. 
\end{itemize}

The AI tool is deemed ``good'' if it is correct for a large enough portion of the test values. 

In general, however, it is difficult to get an insight into the intrinsic decision mechanisms of a typical AI tool; this is especially true for the deep neural networks, which are applied widely today.

In the case of linear Support Vector Machines \cite{Cortes1995}, the decision mechanism is much more transparent, and one can exploit a highly correct SVM for gaining unprecedented scientific information in certain cases \cite{Keresztes2019}. In the present contribution, we show a novel and original method for gaining site-specific amyloid-forming properties of amino acids in hexapeptides and preparing amyloid-forming and non-amyloid forming patterns for the succinct representation of the SVM prediction results for hundreds (cf. Table S1) or even tens of thousands (cf. Table 2) hexapeptides at the same time.

\section*{Methods}

We have introduced the Budapest Amyloid Predictor webserver in the work \cite{Keresztes2020a} by applying linear Support Vector Machines as the underlying prediction tool \cite{Cortes1995}, and the Waltz dataset \cite{Beerten2015, Louros2020} for training and testing purposes. The Waltz dataset consists of 1415 hexapeptides, from which 514 peptides are experimentally labeled as ``amyloidogenic'' and 901 hexapeptides as ``non-amyloidogenic''. The Budapest Amyloid Predictor \url{https://pitgroup.org/bap} was constructed as follows:

\begin{itemize}
\item[(i)] Each amino acid from the 20 proteinogenic ones was characterized by a 553-dimensional vector, corresponding to its physico-chemical properties published in AAindex \cite{Kawashima2008}. Therefore, a hexapeptide was represented by a length 6 $\times$ 553 = 3318 vector $z$. We note that this highly redundant representation has given somewhat better predictions than more concise ones \cite{Keresztes2020a} and has not caused any difficulties in what follows.

\item[(ii)] By applying a quadratic programming algorithm for SVM computation from the SciKit-learn Python library \cite{scikit-learn}, we have computed a vector $w$ and a scalar $b$ such that if the sign of $w\cdot z+b$  is positive, then the prediction is ``amyloidogenic'', otherwise it is ``non-amyloidogenic'', with $84\%$ accuracy, for any vector $z$, representing a hexapeptide.

\item[(iii)] One can write the dot product $w\cdot z$, with $\ell=553$, as
$$w\cdot z= \sum_{i=1}^{6\ell}w_iz_i=\sum_{j=1}^6  \ \ \sum_{i=(j-1)\ell+1}^{j\ell}w_iz_i. \eqno{(1)}  $$
For any given $j=1,2,\ldots,6$ the $\ell$ $z_i'$s are determined by the amino acid at position $j$ at the hexapeptide. This means that we have only $6 \times 20=120$ second sums in (1) (for six positions and 20 amino acids), and these 120 values can be pre-computed. Table 1 of \cite{Keresztes2020a} lists these pre-computed values. Since we need the same table in the present work, we include it also as Table 1 here. 

\item[(iv)] Table 1 makes it possible to decide if a hexapeptide is predicted to be amyloidogenic or not, by ``hand''; for example, to decide if IVIVIV is amyloidogenic or not, we need to add up the numbers, corresponding to I in the first, to V in the second, to I in the third, to V in the fourth, to I in the fifth and to V in the sixth column, that is 
$$-0.06-0.14+0.26+0.14-0.06+0.01=0.15$$
and we need to add to this $w\cdot z$ value the scalar $b=1.083$, which equals to $1.233$, which is a positive number, so IVIVIV is predicted to be amyloidogenic.
\end{itemize}

\begin{table}[H]
	\begin{center}
		\includegraphics[width=8cm]{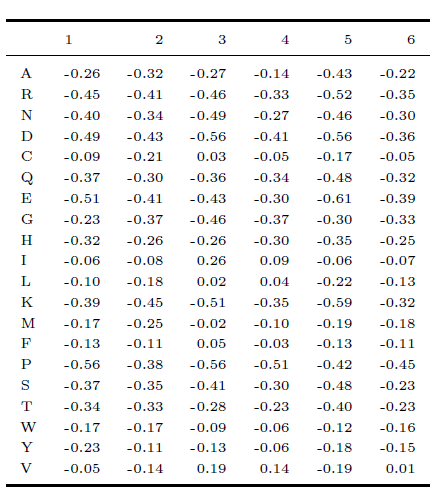}
\caption{ The Amyloid Effect Matrix, constructed from the pre-computed values from equation (1). The rows are corresponding to the amino acids, the columns to the positions. The larger numbers show stronger amyloidogenic properties in the given position. Source: \cite{Keresztes2020a}. In \cite{Keresztes2020a}, by ordering the columns of this table, a position-dependent amyloidogenecity order of amino acids are given in a subsequent table.}
	\end{center}
\end{table}

We refer to Table 1 as the  Amyloid Effect Matrix.

As we have demonstrated in paragraph (iv) above, one can simply make the prediction of the SVM by using the values solely from this matrix. 

From now on, we would like to exploit the Amyloid Effect Matrix for finding succinct descriptions of amyloidogenic and non-amyloidogenic patterns among the 64 million possible hexapeptides. 

\subsection*{Patterns of Amyloidicity}

Here we would like to find very characteristic positions and substitutions, which already assure us that all the hexapeptides fitting those patterns are homogenously either amyloidogenic or non-amyloidogenic. Let us see an example:

\begin{example} 
Let us fix the amino acid proline (P stands for) at positions 3 and 4 and leave all the other four positions free: let us consider the pattern
$$xxPPxx.$$

We state that for all (independent) substitutions for $x$'s, the Budapest Amyloid Predictor (abbreviated as BAP) says that the hexapeptide is NOT amyloid. Since we have four $x$'s, the pattern $xxPPxx$ describes exactly $20^4=160 000$ hexapeptides; so we state that not one of these 160 000 hexapeptides is predicted to be amyloidogenic. 

It is very easy to verify this statement from Table 1. The values corresponding to P's in the third and in the fourth positions (-0.56 and -0.51) add up to -1.07. Now, even if we take the largest values of columns 1, 2, 5, and 6, that is, -0.05, -0.08, -0.06, and 0.01, respectively, their sum is -1.25, and adding  $b=1.083$ to this value, we would still have a result to be a negative number. That is, even the largest values from columns 1,2,5, and 6 could not overweight the large negative sum of -1.07 of the two consecutive proline residues in positions 3 and 4. This means that all hexapeptides, fitting to the pattern of $xxPPxx$, are predicted to be non-amyloids by BAP. 
\end{example}

\begin{example}
Similarly, one can also find amyloid patterns. For example, we state that all the 400 ($= 20 \times 20$) hexapeptides, fitting to the pattern FxFLWx are predicted to be amyloids. One can easily verify this statement from Table 1. The F in position 1 adds -0.13, in position 3 adds 0.05, L in position 4 adds 0.04, and W in position 5 contributes -0.12, their sum is -0.16. Now, if we take the smallest values from columns 2 and 6, that is, -0.45 and -0.45, and add  $b=1.083$ to their sum, we will get -0.16-0.45-0.45+1.083=0.023, that is, a positive number, so independently from the choice of the x's, FxFLWx is predicted to be an amyloid-forming hexapeptide.
\end{example}

\subsection*{Minimal patterns}

In what follows, we will find all the minimal patterns of amyloidicity and non-amyloidicity. These minimal patterns are the most concise representations of the amyloid-forming rules of the BAP predictor.

Here the ``minimal'' word means that we cannot decrease the number of the fixed amino acids without invalidating the rule. Our goal is to find the patterns with the minimum number of amino acids fixed. From such minimal patterns, one can easily generate valid but non-minimal ones; for example, the {\tt  xxPPxx} pattern is predicted to be non-amyloidogenic for any substitutions of x's. Therefore, e.g., {\tt WxPPxx}, or {\tt VIPPxx} are also non-amyloid patterns for any substitutions for $x$,  but they are not minimal. It is easy to see by observing Table 1 that neither {\tt xxxPxx}, nor {\tt xxPxxx} are valid non-amyloid patterns, so {\tt xxPPxx} is a minimal pattern.

\subsection*{Finding all minimal patterns}

Our goal is to find every hexapeptide pattern, both the amyloidogenic and the non-amyloidogenic ones, as predicted by BAP.

Finding these patterns is straightforward using the Amyloid Effect Matrix (Table 1). Suppose that we intend to generate the minimal amyloid indicating patterns. Finding the non-amyloid patterns is a similar procedure.

Verifying whether a pattern is a valid amyloid indicator is easy: we need to generalize the steps done in the examples: we substitute the minimal amyloid effective amino acids on the free positions (denoted by $s$), and check its score. If the score is already positive, then this least amyloidogenic hexapeptide is already amyloid, then every other hexapeptide from this space is amyloid too.

Finding the rules for hexapeptides could be done by exhaustive search. Let say we want to find all the rules with $k$ fixed amino acids, where $k$ is between 1 and 6. In what follows, we call the core of the rule is the number of fixed amino acids (e.g., the core of rule xxPPxx is 2). The positions $s$ will be referred to as free positions.

For finding all the rules with core $k$, our approach is 

\begin{itemize}
\item[(i)] generating all the $ 6 \choose k$ index subsets;

\item[(ii)] for each index subset, we generate all the $20^k$ rule candidates by assigning all the possible amino acids to the $k$ core positions;

\item[(iii)] verify the validity of the pattern by using check each of them as already described;

\end{itemize}

We remark that this exact exhaustive search is not fast computationally, but it perfectly works for hexapeptides: The number of verifications is $$\sum_{k=1}^{k=6}{ {6 \choose k} 20^k} = 21^6-1$$  less than 86 million, its running time is several hours in today's low-end computers.

The amyloid patterns are listed in Table S1 in the Supporting Material, while the non-amyloid patterns are in Table 2 in the next section.

\subsection*{The case of restricted amino acid classes}

Amino acids are frequently characterized and classified by their chemical properties, like polarity, non-polarity, hydrophobicity, hydrophilicity, etc. If we want to find patterns of amyloidicity when for the free positions, denoted by $x$, one can choose substitutions only from a given restricted class, then one can have stronger, more specific patterns than in the general case, when $x$ can be substituted by any of the 20 amino acids. 

Finding those patterns in the restricted classes can be done analogously to the general case: the minimum values of Table 1 from the given class need to be considered.

\subsection*{Statistical analysis}

We refer to the work \cite{Keresztes2020a} for the statistical accuracy estimations of the Budapest Amyloid Predictor. There we have shown that the predictor has ACC = 0.84, TPR=0.75, TNR=0.9, PPV=0.8, NPV=0.86  (that is, accuracy, true positive ratio, true negative ratio, positive predictive value, negative predictive value, resp.) Figure 1 of \cite{Keresztes2020a} also gives ROC (Receiver Operating Characteristics) curve of the tool, with AUC (Area Under Curve) value is 0.89.

\section*{Discussion and Results}

Figure 1 visualizes the substitutions into a non-amyloid and an amyloid-pattern. 

We have successfully identified all the 5531 minimal amyloid patterns (Table S1) using BAP. For example, xxIYCI, IFIYxx, or CxVVxC are amyloid patterns from Table S1.   

\begin{figure}[H]
	\centering
	\includegraphics[scale=0.60]{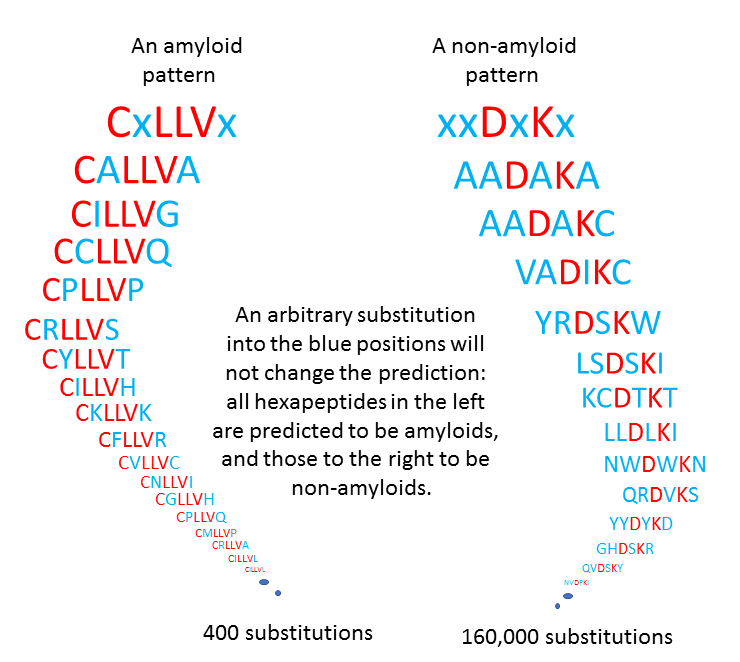}
	\caption{Examples for amyloid- and non-amyloid patterns.}
	\end{figure}

We have found that almost all amyloidogenic hexapeptide patterns contain valine (V) and/or isoleucine (I) residues, both of branched and hydrophobic sidechains. Among the 5531 patterns identified (Table S1) only 16 patterns are free of both V and I:

{ \tt
\begin{verbatim}
CxFLWx, CxFLFx, CxCLWx, CxCLFx, CxLLWx, CxLLFx, LxFLWx, LxFLFx,  
LxCLWx, LxCLFx, LxLLWx, LxLLFx, FxFLWx, FxFLFx, FxCLWx, CxFLCx.
\end{verbatim}}

Furthermore leucine, L, the third branched and hydrophobic side chains is represented one or more times in the above listed 15 patterns. In conclusion, V, I and L residues make hexapeptides intrinsically amyloidogenic.

We remark that no amyloid pattern with three free positions (i.e., x) exists.

Table 2 lists the 24 non-amyloid patterns, each with four free positions. No other non-amyloid patterns exist with four free positions, and no non-amyloid pattern exists with five free positions.

It is interesting to note that from the 24 non-amyloid patterns of Table 2, 15 contain proline in one or two fixed positions. From the 9 remainders, 4 contain E-K or D-K pairs, which can form possibly stabilizing salt bridges \cite{Nagy2010}.

\begin{table}
\centering
{\tt
\begin{tabular}{cccccc} 
PxPxxx & PxDxxx & xxPPxx & xxPDxx & xxPGxx & xxPKxx \\
xxPQxx & xxDPxx & xxDDxx & xxDGxx & xxDKxx & xxDQxx \\
xxKPxx & xxKDxx & xxNPxx & xxGPxx & xxRPxx & xxPxEx \\
xxPxKx & xxPxDx & xxDxEx & xxDxKx & xxDxDx & xxKxEx 
\end{tabular}
}
\caption{The list of all non-amyloid patterns with four free positions; it contains 24 patterns. Note that each pattern describes $20^4=160 000$ hexapeptides succinctly, all of which are predicted to be non-amyloids by the Budapest Amyloid Predictor \cite{Keresztes2020a}. From the 24 patterns, only nine do not contain proline in a fixed position. Note also the possibly stabilizing salt-bridge forming \cite{Nagy2010} amino acid pairs (E-K, D-K) in several patterns without the fixed proline. }
\end{table}

\subsection*{Results for amino acid subsets}

In this subsection, we find amyloid patterns when the $x$ positions can be substituted only by the members of some specific amino acid classes. 
The amino acid classes we examine are small non-polar, hydrophobic, and polar amino acids, as classified by \cite{Lesk}, and listed in the second column of Table 3.

\begin{table}[H]
    \centering
    \small
    \setlength\tabcolsep{3pt}
    \begin{tabular}{llcr}

Class name &       Class elements  &  No. of free  &  No. of  \\
& listing & positions & patterns  \\
\midrule
    small non-polar &                GAST & 3 &  411 \\
       hydrophobic &            CVLIMPFYW & 3 &   43 \\
             polar &              DENQHKR & 3 &    4 \\
\midrule
 hydrophobic-\{P\} &             CVLIMFYW & 5 &   38 \\
 amino acids-\{P\} &  QFYESNCDMLIAHGWRKVT & 3 &    4 \\
\bottomrule
\end{tabular}
    \caption{The amino acid subsets examined. The classification of the residues in the first three rows are by \cite{Lesk}. The last two rows correspond to the classes where we left out proline, a well-known structure-breaker from the hydrophobic set or from all the amino acids. The third column shows the number of free positions we get in the special substitutions, and the fourth column shows the number of patterns found for these special substitutions for ``x''.}
\end{table}

When the substitutions to the free positions, denoted by x, can be done only from special subsets, listed in Table 3, we can get amyloid-rules with 3 free positions, in contrast with the unrestricted case, when our rules have 2 free positions (Table S1). 

When x is allowed to be substituted from the small non-polar set, then 411 patterns can be found with three free positions, for example, {\tt VIIxxx IIIxxx VxIVxx VxIIxx VxILxx VxIFxx
VxICxx VxIWxx VxVVxx VxVIxx}. All the existing 411 patterns are listed in Table S2 in the supporting material. No such pattern exists with 4 free positions.

If x is chosen from the hydrophobic set, then 43 patterns exist with three free positions, listed in Table 4. No such pattern exists with 4 free positions.

\begin{table}[H]
\centering
{\tt
\begin{tabular}{llllllll}
VxIVxx & VxIIxx & VxILxx & VxIFxx & VxVVxx & VxVIxx & VxVLxx & IxIVxx  \\
IxIIxx & IxILxx & IxVVxx & IxVIxx & CxIVxx & CxIIxx & CxILxx & CxVVxx \\
CxVIxx & LxIVxx & LxIIxx & LxILxx & LxVVxx & LxVIxx & FxIVxx & FxIIxx \\
FxVVxx & MxIVxx & MxIIxx & MxVVxx & WxIVxx & WxIIxx & GxIVxx & YxIVxx \\
xxIVIx & xxIVxV & xxIVxC & xxIVxI & xxIVxF & xxIIxV & xxIIxC & xxILxV \\
xxVVxV & xxVVxC & xxVIxV &        &        &        &        &        \\
\end{tabular}
}
\caption{The list of all the 43 amyloidogenic patterns with three free positions when x is hydrophobic, i.e., chosen from CVLIMPFYW. Each pattern describes $9^3=729$ hexapeptides.}
\end{table}

When x is chosen from polar amino acids, then the only four patterns with three free positions are {\tt xxIVIx xxIVWx xxIIIx xxVVIx}.

We note that no pattern exists in these three cases without V and I amino acids; that is, all of the patterns in these three restricted substitutions contain either valine or isoleucine in fixed positions.

If proline is not allowed to be substituted for any x, but otherwise the remaining 19 amino acids can be chosen for the x positions, then we have exactly four amyloid patterns with three x positions: xxIVIx xxIVWx xxIIIx xxVVIx; note that without the restriction to proline, no amyloid pattern exists with three free positions. 

These four patterns are exactly the same as in the case of polar residue substitutions, but the set of hexapeptides they represent differ: in the case of polar substitutions, each of the four patterns represent $7^3=343$ hexapeptides, while for the non-proline substitutions $19^3=6859$ hexamers.

If $x$ could be chosen from hydrophobic amino acids, except proline, the ``structure breaker'' then we have the ``largest'' patterns of amyloidicity: 38 patterns exist with just one fixed position, listed in Table 5. Note that each of those patterns describes $8^5=32768$ hexapeptides, such that all of them are predicted to be amyloidogenic. 

\begin{table}
\centering
{\tt
\begin{tabular}{cccccccc} 
Vxxxxx & Ixxxxx & Cxxxxx & Lxxxxx & Fxxxxx & Mxxxxx & Wxxxxx & xIxxxx  \\
xFxxxx & xYxxxx & xVxxxx & xWxxxx & xLxxxx & xCxxxx & xxIxxx & xxVxxx  \\
xxFxxx & xxCxxx & xxLxxx & xxMxxx & xxWxxx & xxxVxx & xxxIxx & xxxLxx \\
xxxFxx & xxxCxx & xxxWxx & xxxYxx & xxxxIx & xxxxWx & xxxxFx & xxxxCx \\
xxxxYx & xxxxxV & xxxxxC & xxxxxI & xxxxxF & xxxxxL &        &         \\
 
\end{tabular}
}
\caption{The list of all the 38 amyloidogenic patterns with five free positions when x is hydrophobic, but cannot be proline, i.e., chosen from CVLIMFYW. Each pattern describes $8^5=32768$ hexapeptides.}
\end{table}

\section*{Conclusions} Here, we established patterns of amyloidicity  and non-amyloidicity in the case of hexapeptides, based on a Support Vector Machine--based predictor, available at \url{https://pitgroup.org/bap}. To our knowledge, no machine learning tool was used before to derive succinct chemical knowledge through simple patterns for deep structural properties. 
	
\section*{Data availability} The Budapest Amyloid Predictor webserver is available freely at \url{https://pitgroup.org/bap}.

\section*{Funding}
LK, ES, BV and VG were partially supported by the  NKFI-127909 grant of the National Research, Development and Innovation Office of Hungary.  LK, ES, AP, VF and VG were partially supported by the ELTE Thematic Excellence Programme (Szint+) subsidized by the Hungarian Ministry for Innovation and Technology.

\section*{Author Contribution} AP, VF and VG have initiated the study and evaluated results, LK and ES constructed the SVM for the prediction, BV constructed the webserver, VG has overseen the work and wrote the paper. AP, VF and VG secured funding.

\section*{Conflicting interest} The authors declare no conflicting interests.

%\bibliography{v:/vince/CIKKEK/medl}

\begin{thebibliography}{22}
\providecommand{\natexlab}[1]{#1}
\providecommand{\url}[1]{\texttt{#1}}
\expandafter\ifx\csname urlstyle\endcsname\relax
  \providecommand{\doi}[1]{doi: #1}\else
  \providecommand{\doi}{doi: \begingroup \urlstyle{rm}\Url}\fi

\bibitem[Michiels et~al.(2020)Michiels, Roose, Gallardo, Khodaparast,
  Khodaparast, van~der Kant, Siemons, Houben, Ramakers, Wilkinson, Guerreiro,
  Louros, Kaptein, Ibanez, Smet, Baatsen, Liu, Vorberg, Bormans, Neyts,
  Saelens, Rousseau, and Schymkowitz]{Michiels2020}
Emiel Michiels, Kenny Roose, Rodrigo Gallardo, Ladan Khodaparast, Laleh
  Khodaparast, Rob van~der Kant, Maxime Siemons, Bert Houben, Meine Ramakers,
  Hannah Wilkinson, Patricia Guerreiro, Nikolaos Louros, Suzanne J~F Kaptein,
  Lorena~Itati Ibanez, Anouk Smet, Pieter Baatsen, Shu Liu, Ina Vorberg, Guy
  Bormans, Johan Neyts, Xavier Saelens, Frederic Rousseau, and Joost
  Schymkowitz.
\newblock Reverse engineering synthetic antiviral amyloids.
\newblock \emph{Nature Communications}, 11:\penalty0 2832, June 2020.
\newblock ISSN 2041-1723.
\newblock \doi{10.1038/s41467-020-16721-8}.

\bibitem[Gillmore et~al.(0)Gillmore, Gane, Taubel, Kao, Fontana, Maitland,
  Seitzer, O'Connell, Walsh, Wood, Phillips, Xu, Amaral, Boyd, Cehelsky, McKee,
  Schiermeier, Harari, Murphy, Kyratsous, Zambrowicz, Soltys, Gutstein,
  Leonard, Sepp-Lorenzino, and Lebwohl]{Gillmore2021}
Julian~D. Gillmore, Ed~Gane, Jorg Taubel, Justin Kao, Marianna Fontana,
  Michael~L. Maitland, Jessica Seitzer, Daniel O'Connell, Kathryn~R. Walsh,
  Kristy Wood, Jonathan Phillips, Yuanxin Xu, Adam Amaral, Adam~P. Boyd,
  Jeffrey~E. Cehelsky, Mark~D. McKee, Andrew Schiermeier, Olivier Harari,
  Andrew Murphy, Christos~A. Kyratsous, Brian Zambrowicz, Randy Soltys,
  David~E. Gutstein, John Leonard, Laura Sepp-Lorenzino, and David Lebwohl.
\newblock Crispr-cas9 in vivo gene editing for transthyretin amyloidosis.
\newblock \emph{New England Journal of Medicine}, 0\penalty0 (0):\penalty0
  null, 0.
\newblock \doi{10.1056/NEJMoa2107454}.
\newblock URL \url{https://doi.org/10.1056/NEJMoa2107454}.

\bibitem[Horv{\'a}th et~al.(2019)Horv{\'a}th, Menyh{\'a}rd, and
  Perczel]{Horvath2019}
D{\'a}niel Horv{\'a}th, D{\'o}ra~K Menyh{\'a}rd, and Andr{\'a}s Perczel.
\newblock Protein aggregation in a nutshell: The splendid molecular
  architecture of the dreaded amyloid fibrils.
\newblock \emph{Current Protein and Peptide Science}, 20\penalty0
  (11):\penalty0 1077--1088, 2019.

\bibitem[Taricska et~al.(2020)Taricska, Horvath, Menyhard, Akontz-Kiss, Noji,
  So, Goto, Fujiwara, and Perczel]{Taricska2020}
Nora Taricska, Daniel Horvath, Dora~K Menyhard, Hanna Akontz-Kiss, Masahiro
  Noji, Masatomo So, Yuji Goto, Toshimichi Fujiwara, and Andras Perczel.
\newblock The route from the folded to the amyloid state: Exploring the
  potential energy surface of a drug-like miniprotein.
\newblock \emph{Chemistry A European Journal}, 26:\penalty0 1968--1978,
  February 2020.
\newblock ISSN 1521-3765.
\newblock \doi{10.1002/chem.201903826}.
\newblock URL \url{https://doi.org/10.1002/chem.201903826}.

\bibitem[Tak{\'a}cs et~al.(2019)Tak{\'a}cs, Varga, and Grolmusz]{Takacs2019}
Krist{\'o}f Tak{\'a}cs, B{\'a}lint Varga, and Vince Grolmusz.
\newblock Pdb \_amyloid: an extended live amyloid structure list from the
  {PDB}.
\newblock \emph{FEBS Open Bio}, 9\penalty0 (1):\penalty0 185--190, 2019.

\bibitem[Takacs and Grolmusz(2021)]{Takacs2020}
Kristof Takacs and Vince Grolmusz.
\newblock On the border of the amyloidogenic sequences: Prefix analysis of the
  parallel beta sheets in the {PDB\_Amyloid} collection.
\newblock \emph{Journal of Integrative Bioinformatics}, 2021.
\newblock URL \url{https://doi.org/10.1515/jib-2020-0043}.

\bibitem[Maji et~al.(2009)Maji, Perrin, Sawaya, Jessberger, Vadodaria, Rissman,
  Singru, Nilsson, Simon, Schubert, et~al.]{Maji2009}
Samir~K Maji, Marilyn~H Perrin, Michael~R Sawaya, Sebastian Jessberger, Krishna
  Vadodaria, Robert~A Rissman, Praful~S Singru, K~Peter~R Nilsson, Rozalyn
  Simon, David Schubert, et~al.
\newblock Functional amyloids as natural storage of peptide hormones in
  pituitary secretory granules.
\newblock \emph{Science}, 325\penalty0 (5938):\penalty0 328--332, 2009.

\bibitem[Soto et~al.(2006)Soto, Estrada, and Castilla]{Soto2006}
Claudio Soto, Lisbell Estrada, and Joaqu{\'\i}n Castilla.
\newblock Amyloids, prions and the inherent infectious nature of misfolded
  protein aggregates.
\newblock \emph{Trends in Biochemical Sciences}, 31\penalty0 (3):\penalty0
  150--155, 2006.

\bibitem[Familia et~al.(2015)Familia, Dennison, Quintas, and
  Phoenix]{Familia2015}
Carlos Familia, Sarah~R Dennison, Alexandre Quintas, and David~A Phoenix.
\newblock Prediction of peptide and protein propensity for amyloid formation.
\newblock \emph{PLoS ONE}, 10:\penalty0 e0134679, 2015.
\newblock ISSN 1932-6203.
\newblock \doi{10.1371/journal.pone.0134679}.

\bibitem[Tartaglia and Vendruscolo(2008)]{Tartaglia2008a}
Gian~Gaetano Tartaglia and Michele Vendruscolo.
\newblock The zyggregator method for predicting protein aggregation
  propensities.
\newblock \emph{Chemical Society reviews}, 37:\penalty0 1395--1401, July 2008.
\newblock ISSN 0306-0012.
\newblock \doi{10.1039/b706784b}.

\bibitem[Conchillo-Sole et~al.(2007)Conchillo-Sole, de~Groot, Aviles, Vendrell,
  Daura, and Ventura]{Conchillo-Sole2007}
Oscar Conchillo-Sole, Natalia~S de~Groot, Francesc~X Aviles, Josep Vendrell,
  Xavier Daura, and Salvador Ventura.
\newblock Aggrescan: a server for the prediction and evaluation of "hot spots"
  of aggregation in polypeptides.
\newblock \emph{BMC Bioinformatics}, 8:\penalty0 65, February 2007.
\newblock ISSN 1471-2105.
\newblock \doi{10.1186/1471-2105-8-65}.

\bibitem[Kim et~al.(2009)Kim, Choi, Lee, Welsh, and Yoon]{Kim2009a}
Changsik Kim, Jiwon Choi, Seong~Joon Lee, William~J Welsh, and Sukjoon Yoon.
\newblock Netcssp: web application for predicting chameleon sequences and
  amyloid fibril formation.
\newblock \emph{Nucleic acids research}, 37:\penalty0 W469--W473, July 2009.
\newblock ISSN 1362-4962.
\newblock \doi{10.1093/nar/gkp351}.

\bibitem[Santos et~al.(2020)Santos, Pujols, Pallares, Iglesias, and
  Ventura]{Santos2020}
Jaime Santos, Jordi Pujols, Irantzu Pallares, Valentin Iglesias, and Salvador
  Ventura.
\newblock Computational prediction of protein aggregation: Advances in
  proteomics, conformation-specific algorithms and biotechnological
  applications.
\newblock \emph{Computational and Structural Biotechnology Journal},
  18:\penalty0 1403--1413, 2020.
\newblock ISSN 2001-0370.
\newblock \doi{10.1016/j.csbj.2020.05.026}.

\bibitem[Keresztes et~al.()Keresztes, Szogi, Varga, Farkas, Perczel, and
  Grolmusz]{Keresztes2020a}
Laszlo Keresztes, Evelin Szogi, Balint Varga, Viktor Farkas, Andras Perczel,
  and Vince Grolmusz.
\newblock The {B}udapest {A}myloid {P}redictor and its applications.
\newblock \emph{Biomolecules}, 11\penalty0 (4):\penalty0 500.
\newblock URL \url{https://doi.org/10.3390/biom11040500}.

\bibitem[Cortes and Vapnik(1995)]{Cortes1995}
Corinna Cortes and Vladimir Vapnik.
\newblock Support-vector networks.
\newblock \emph{Machine Learning}, 20\penalty0 (3):\penalty0 273--297, 1995.

\bibitem[Beerten et~al.(2015)Beerten, Van~Durme, Gallardo, Capriotti, Serpell,
  Rousseau, and Schymkowitz]{Beerten2015}
Jacinte Beerten, Joost Van~Durme, Rodrigo Gallardo, Emidio Capriotti, Louise
  Serpell, Frederic Rousseau, and Joost Schymkowitz.
\newblock {WALTZ-DB}: a benchmark database of amyloidogenic hexapeptides.
\newblock \emph{Bioinformatics (Oxford, England)}, 31:\penalty0 1698--1700, May
  2015.
\newblock ISSN 1367-4811.
\newblock \doi{10.1093/bioinformatics/btv027}.

\bibitem[Louros et~al.(2020)Louros, Konstantoulea, De~Vleeschouwer, Ramakers,
  Schymkowitz, and Rousseau]{Louros2020}
Nikolaos Louros, Katerina Konstantoulea, Matthias De~Vleeschouwer, Meine
  Ramakers, Joost Schymkowitz, and Frederic Rousseau.
\newblock {WALTZ-DB} 2.0: an updated database containing structural information
  of experimentally determined amyloid-forming peptides.
\newblock \emph{Nucleic Acids Research}, 48:\penalty0 D389--D393, January 2020.
\newblock ISSN 1362-4962.
\newblock \doi{10.1093/nar/gkz758}.

\bibitem[Kawashima et~al.(2008)Kawashima, Pokarowski, Pokarowska, Kolinski,
  Katayama, and Kanehisa]{Kawashima2008}
Shuichi Kawashima, Piotr Pokarowski, Maria Pokarowska, Andrzej Kolinski,
  Toshiaki Katayama, and Minoru Kanehisa.
\newblock Aaindex: amino acid index database, progress report 2008.
\newblock \emph{Nucleic Acids Research}, 36:\penalty0 D202--D205, January 2008.
\newblock ISSN 1362-4962.
\newblock \doi{10.1093/nar/gkm998}.

\bibitem[Keresztes et~al.(2021)Keresztes, Szogi, Varga, and
  Grolmusz]{Keresztes2019}
Laszlo Keresztes, Evelin Szogi, Balint Varga, and Vince Grolmusz.
\newblock Identifying super-feminine, super-masculine and sex-defining
  connections in the human braingraph.
\newblock \emph{Cognitive Neurodynamics}, 15\penalty0 (6):\penalty0 949--959,
  2021.
\newblock URL \url{https://doi.org/10.1007/s11571-021-09687-w}.

\bibitem[Pedregosa et~al.(2011)Pedregosa, Varoquaux, Gramfort, Michel, Thirion,
  Grisel, Blondel, Prettenhofer, Weiss, Dubourg, Vanderplas, Passos,
  Cournapeau, Brucher, Perrot, and Duchesnay]{scikit-learn}
F.~Pedregosa, G.~Varoquaux, A.~Gramfort, V.~Michel, B.~Thirion, O.~Grisel,
  M.~Blondel, P.~Prettenhofer, R.~Weiss, V.~Dubourg, J.~Vanderplas, A.~Passos,
  D.~Cournapeau, M.~Brucher, M.~Perrot, and E.~Duchesnay.
\newblock Scikit-learn: Machine learning in {P}ython.
\newblock \emph{Journal of Machine Learning Research}, 12:\penalty0 2825--2830,
  2011.

\bibitem[Nagy and Erhardt(2010)]{Nagy2010}
Peter~I Nagy and Paul~W Erhardt.
\newblock Theoretical studies of salt-bridge formation by amino acid side
  chains in low and medium polarity environments.
\newblock \emph{The Journal of Physical Chemistry B}, 114\penalty0
  (49):\penalty0 16436--16442, 2010.
\newblock URL \url{https://doi.org/10.1021/jp103313s}.

\bibitem[Lesk(2007)]{Lesk}
Arthur~M Lesk.
\newblock \emph{Introduction to Bioinformatics}.
\newblock Oxford: Oxford University Press, 2007.

\end{thebibliography}
%\bibliographystyle{unsrtnat}

\section*{Supporting Material}

\section*{Table S1: Amyloid patterns in hexapeptides}

In the table we list all the existing minimal patterns of amyloid-forming hexapeptides, computed from the SVM model of the Budapest Amyloid Predictor. Each pattern in the Table contains two free ``x'' positions, where any amino acids can be substituted, and the resulting hexapeptide is predicted to be amyloidogenic.  Since the two free positions can be substituted $20 \times 20 = 400$ ways, each of the 5531 patterns in the Table describes 400 hexapeptides succinctly. 

{\scriptsize \tt

\begin{longtable}{llllllllll}
\toprule
     0 &      1 &      2 &      3 &      4 &      5 &      6 &      7 &      8 &      9 \\
\midrule
VIIVxx & VIIIxx & VIILxx & VIIFxx & VIICxx & VIIWxx & VIIYxx & VIIMxx & VIIAxx & VIVVxx \\
VIVIxx & VIVLxx & VIVFxx & VIVCxx & VIVWxx & VIVYxx & VIFVxx & VIFIxx & VICVxx & VICIxx \\
VILVxx & VILIxx & VIMVxx & VFIVxx & VFIIxx & VFILxx & VFIFxx & VFICxx & VFIWxx & VFIYxx \\
VFIMxx & VFVVxx & VFVIxx & VFVLxx & VFVFxx & VFVCxx & VFFVxx & VFFIxx & VFCVxx & VFLVxx \\
VYIVxx & VYIIxx & VYILxx & VYIFxx & VYICxx & VYIWxx & VYIYxx & VYIMxx & VYVVxx & VYVIxx \\
VYVLxx & VYVFxx & VYVCxx & VYFVxx & VYFIxx & VYCVxx & VYLVxx & VVIVxx & VVIIxx & VVILxx \\
VVIFxx & VVICxx & VVIWxx & VVIYxx & VVVVxx & VVVIxx & VVVLxx & VVVFxx & VVFVxx & VVCVxx \\
\bottomrule
\end{longtable}
}

\newpage

\section*{Table S2: Amyloid patterns for restricted substitutions with small, non-polar residues}

In the table we list all the existing minimal patterns of amyloid-forming hexapeptides, computed from the SVM model of the Budapest Amyloid Predictor, when the substitutions for the positions, denoted by ``x'' are allowed by small, non-polar residues, namely G, A, S and T. The table contains 411 patterns.

{\scriptsize \tt

\begin{longtable}{llllllllll}
\toprule
     0 &      1 &      2 &      3 &      4 &      5 &      6 &      7 &      8 &      9 \\
\midrule
VIIxxx & IIIxxx & VxIVxx & VxIIxx & VxILxx & VxIFxx & VxICxx & VxIWxx & VxIYxx & VxIMxx \\
VxVVxx & VxVIxx & VxVLxx & VxVFxx & VxFVxx & VxCVxx & VxLVxx & IxIVxx & IxIIxx & IxILxx \\
IxIFxx & IxICxx & IxIWxx & IxIYxx & IxVVxx & IxVIxx & IxVLxx & IxVFxx & IxFVxx & IxCVxx \\
IxLVxx & CxIVxx & CxIIxx & CxILxx & CxIFxx & CxICxx & CxIWxx & CxIYxx & CxVVxx & CxVIxx \\
CxVLxx & CxFVxx & LxIVxx & LxIIxx & LxILxx & LxIFxx & LxICxx & LxIWxx & LxVVxx & LxVIxx \\
LxVLxx & FxIVxx & FxIIxx & FxILxx & FxVVxx & FxVIxx & FxVLxx & MxIVxx & MxIIxx & MxILxx \\
MxVVxx & MxVIxx & WxIVxx & WxIIxx & WxILxx & WxVVxx & WxVIxx & GxIVxx & GxIIxx & GxVVxx \\
\bottomrule
\end{longtable}
}

\end{document}